# The Blind Watchmaker Network: Scale-Freeness and Evolution

Petter Minnhagen[1,2], Sebastian Bernhardsson[1,2]*

1 Department of Theoretical Physics, Umeå University, Umeå, Sweden, 2 Center for Models of Life, Copenhagen, Denmark

**Abstract**

It is suggested that the degree distribution for networks of the cell-metabolism for simple organisms reflects a ubiquitous randomness. This implies that natural selection has exerted no or very little pressure on the network degree distribution during evolution. The corresponding random network, here termed *the blind watchmaker network* has a power-law degree distribution with an exponent $\gamma \geq 2$. It is random with respect to a complete set of network states characterized by a description of which links are attached to a node as well as a time-ordering of these links. No a priory assumption of any growth mechanism or evolution process is made. It is found that the degree distribution of the blind watchmaker network agrees very precisely with that of the metabolic networks. This implies that the evolutionary pathway of the cell-metabolism, when projected onto a metabolic network representation, has remained statistically random with respect to a complete set of network states. This suggests that even a biological system, which due to natural selection has developed an enormous specificity like the cellular metabolism, nevertheless can, at the same time, display well defined characteristics emanating from the ubiquitous inherent random element of Darwinian evolution. The fact that also completely random networks may have scale-free node distributions gives a new perspective on the origin of scale-free networks in general.

**Funding:** This work was supported by the Swedish Research Council through contract 50412501.

**Competing Interests:** The authors have declared that no competing interests exist.

* E-mail: sebbeb@tp.umu.se

## Introduction

A network is a representation of who or what is connected to, or influenced by, whom or what. To characterize the structure of a network, researchers often measure its degree distribution $N(k)$, the number of nodes with $k$ links attached. Numerous studies have found that real-world networks often have very broad degree distributions for larger $k$, $N(k) \sim k^{-\gamma}$; These fat tailed distributions approximate a power law in their structure [1][2][3][4][5]. They are also called scale-free networks because a power-law tail indicates the lack of an intrinsic characteristic degree size. Biological networks are particularly interesting because the structure of these networks have, directly or indirectly, arisen through the process of evolution by natural selection. These networks have been constructed as if by a blind watchmaker, through the interplay between a random stochastic evolution and a natural selection process [6]. So what can we learn from the observation that a biological network such as the network of the metabolism of a cell has a power law distribution [7]? In order to answer this question we define and investigate a blind watchmaker network. We demonstrate that the empirically observed degree distributions for networks of the cellular metabolism for simple organisms are good approximations of this random structure. Previous authors have ascribed the scale-free structure of biological networks to various aspects of the evolutionary process: Either the scale-free network structure has been suggested to confer an evolutionary advantage [8][9], or the elementary mechanism for growth of the network has been suggested to generate a priori a scale-free network [10][11][12][13]. Our findings are to the contrary: The close correspondence between the blind watchmaker network and the structure of empirical networks implies that the evolutionary pathway leading to the construction of the cell-metabolism, when projected onto our network representation, is statistically random with respect to a complete set of network states (see Figs. 1 and 2). Thus our findings provide a new perspective on the origin of scale-free network structures [1][2][3][4][5].

The cells living today have evolved into their present forms during some $4 \times 10^9$ years. An important part of the cell is its metabolism which provides the cell with substances necessary for sustaining life. It is a chemical mini-factory usually involving of the order of 500 to 1000 substances in a complex network of metabolic reactions. Some substances like water and $NAD^+$ take part in very many chemical reactions producing new substances either needed in the chemical mini-factory itself to produce other substances, or needed in some other function in the cell, or in fact both [14][15]. Other substances like iron are just needed for some specific purpose in the cell.

One possible metabolic network representation is constructed as follows: Substrates and products in the cell metabolism are nodes. Two nodes are connected if the substance of one is a substrate in a metabolic reaction which produces the substance represented by the other node. This means that each node corresponds to a particular substance and that the links denote its connections to other substances. One characteristic feature of a network is its degree distribution $N(k)$. In the present context this is the number of substances which are connected to precisely $k$ other substances. Figure 3(b) shows the degree distribution for the metabolic network of the E. Coli. bacteria. Here the substance with the most connection is water with 302 connections followed by $NAD^+$ with 141 connections. In Fig. 3(b) these two substances corresponds to the two split off nodes with the largest $k$. The bulk part of the $N(k)$-distribution of E. Coli. is broad and power law like. The straight line in the figure corresponds to $N(k) \sim k^{-\gamma}$ with $\gamma \gtrsim 2$ and illustrates that the node-degree distribution for metabolic networks is broad and power-law like, as was first demonstrated in Ref. [16]. This power law like distribution is even more pronounced when taking an average over many different metabolic networks as shown in Fig. 3(a).

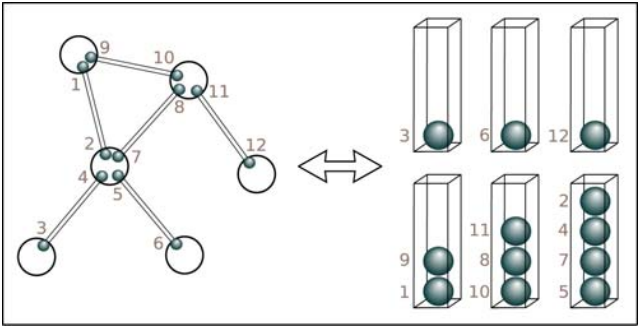

**Figure 1. The mapping between network and the balls-in-boxes model.** The left half shows a network of nodes and links with the link-ends enumerated. The right half shows the equivalent balls-in-boxes model. The vertical position of the balls in a box gives the order in which they arrived into the box, or more generally a ranking.
doi:10.1371/journal.pone.0001690.g001

As the complex metabolic system is projected and reduced into a network, the relevant possibilities also reduce to the corresponding relevant possibilities of the network. A possible state of a network corresponds to a possible way of assembling its parts *i.e.* it includes both a description of what nodes a particular node is connected to and a description of the time order this particular node was connected to these other nodes. The blind watchmaker network is the network which is unbiased with respect to these different assembling possibilities. In the following we will briefly explain what the properties of the blind watchmaker network are and how they are obtained (with more details in supporting information in Text S1).

## Results

We start from a simplified network model, the constrained-balls-in-boxes(CBB)-model [17]. The mapping between the CBB-model and the corresponding network goes as follows: a link is defined by its link-ends such that an enumeration of the links is given by $(1,2),(3,4)\ldots\ldots(M-1,M)$ where the link ends corresponds to balls enumerated by $1,2,\ldots,M$ as is illustrated in Fig. 1: The nodes correspond to boxes and the link-ends to the enumerated balls. The balls in the boxes are given a ranking by the vertical position of the balls in a box. The time-order of connections gives the ranking: earliest connection corresponds to the bottom position and the latest to the top. *The existence of this time-order ranking is crucial in the following.* The point is that the existence of an implicit time-ordering is an un-avoidable consequence of any sequential process. Natural selection is but one example of such a process.

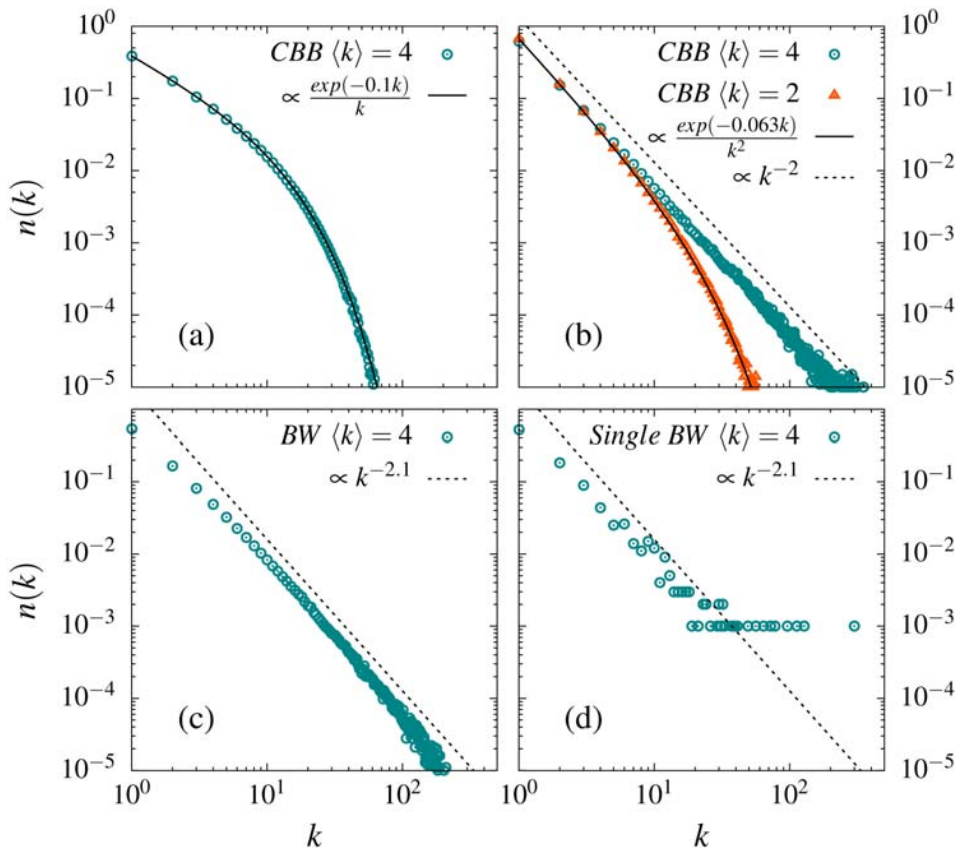

**Figure 2. The most unbiased box(node)-size distribution.** Panels a-d show the most unbiased distributions $n(k)$ obtained by the algorithm method: a) The variational solution $n(k) = A \exp(-bk)/k$ for the statistical states of the CBB-model compared to the average $\langle n(k) \rangle$ solution from the algorithm method in case of $N = 1000$ and $M/N = 4$. The fact that the two solutions agree reflects that the algorithm contains the least possible information compatible with the constraints: b) The same thing for the blind watchmaker version of the CBB-model. In this case the variational solution is $n(k) = A \exp(-bk)/k^2$ and the agreement again reflects that no additional information is contained in the algorithm. The two curves in the panel corresponds to $N = 1000$ and $M/N = 2$ and 4, respectively. As seen the distribution becomes more power law like with increasing ratio $M/N = \langle k \rangle$ : c) The corresponding network solution. The most unbiased solution in this case gives the blind watchmaker network structure. The variational solution cannot be simply obtained for the network case because of the network constraints. By contrast these constraints are easily incorporated into the algorithm method. The panel shows the average solution for $M/N = 4$ and $N = 1000$. The solution has a fat tail power law $n(k) \sim k^{-\gamma}$ with $\gamma \approx 2.1$ : d) A real network is just one representation $n(k)$ and not an average $\langle n(k) \rangle$. The panel shows a single network for $N = 1000$ and $M/N = 4$. A single network always contain large split-off nodes.
doi:10.1371/journal.pone.0001690.g002

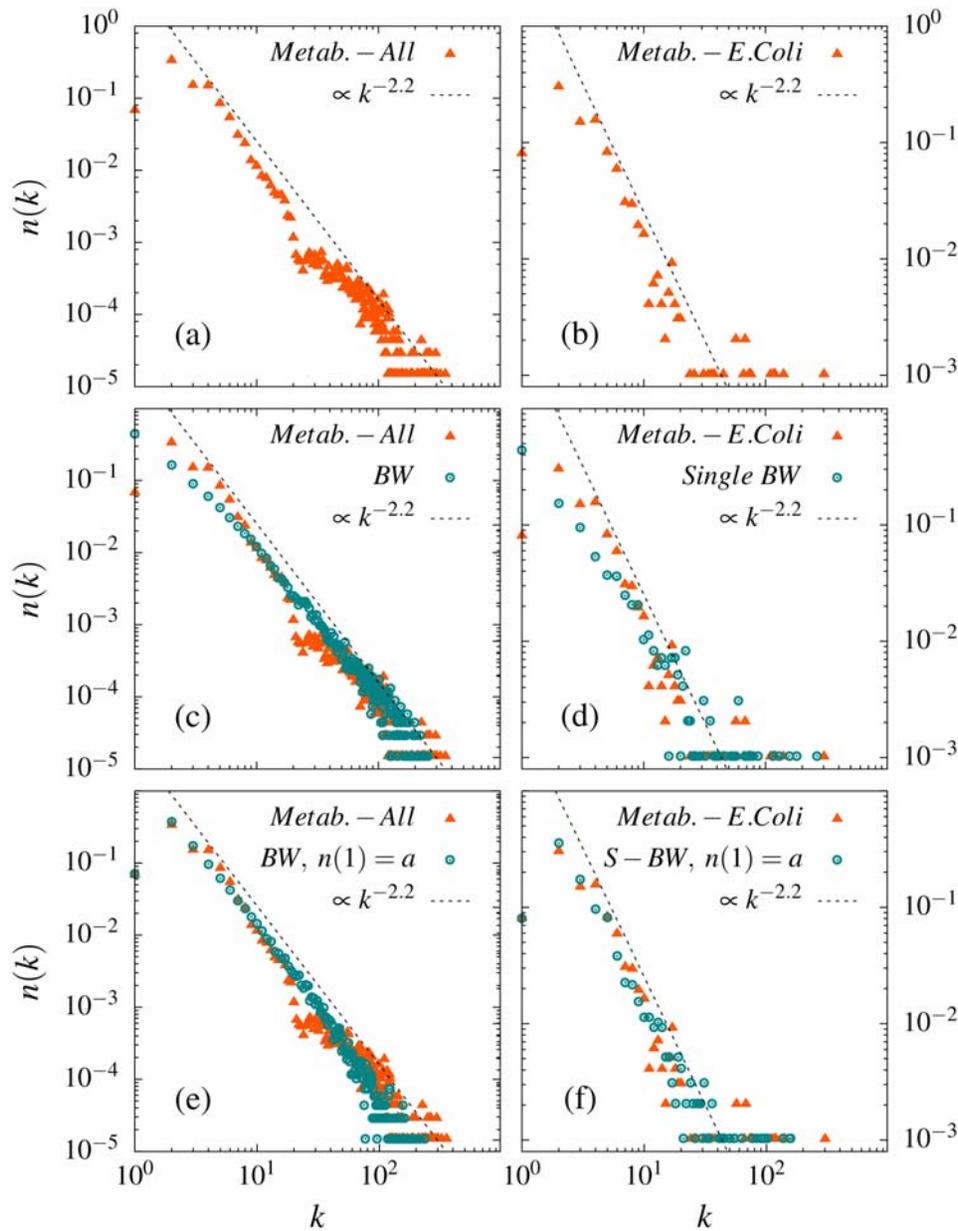

**Figure 3. Metabolic networks and the blind watchmaker network.** The first two panels show the data for real metabolic networks (the data is taken from Ref. [ma03a][ma03b]): Panel a) shows the average distribution $\langle n(k) \rangle$ over 107 metabolic networks. This average distribution has a fat power-law tail $\langle n(k) \rangle \sim k^{-\gamma}$ with $\gamma \approx 2.2$. Note that the nodes with only one link are fewer (by roughly a factor of 5) than the nodes with two links. The average size of these 107 networks is $N \approx 640$ and $M/N \approx 5.35$ b) the specific metabolic network distribution $n(k)$ for the E.Coli bacteria ($N \approx 970$ and $M/N \approx 5.8$). This network has 6 nodes with more than 100 links. Panel c) makes a direct comparison between the 107 metabolic networks in Panel a and 107 blind watchmaker networks with the same $N$ and $M$. The agreement implies a common origin. Panel d) makes the same comparison between the E.Coli network and a single random blind watchmaker network for the same $N$ and $M$. Apart from a general agreement (modulo the larger statistical spread inherent in comparing single networks), the distribution of the split-off nodes are similar. Panel e) compares the average of the metabolic networks with the corresponding average of the blind watchmaker networks *including* a constraint decreasing the abundance of the single link node to the same average number, $n(1)$, as for the metabolic networks ($a=0.07$). The agreement is extraordinary. Panel f) makes the same comparison for the E. Coli. network ($a=0.08$) again with a striking agreement.
doi:10.1371/journal.pone.0001690.g003

For simplicity we here consider undirected links. A network is then an association of link-ends where the associated link-ends form nodes. Note that this means that a node will always contain at least one link-end. This is the origin of "constrained" in the name of the model: a box always contains at least one ball. Obviously, any association of the balls corresponds to a network. However, it is customary to include additional constraints in the definition of a network. The most common for an undirected network are: 1) a network must be connected, 2) only one link between two nodes, 3) no self-loops (which means that the two link-ends of the same link cannot belong to the same node). We will first discuss the model without the additional network constraints. In order to make connection to standard statistical mechanics we will consider the case with a fixed number of balls $M$ and a fixed number of boxes $N$.

The total number of ways you can distribute $M$ balls into $N$ boxes, $\Omega$, is by elementary combinatorics $\Omega = \frac{M!}{\Pi_{k=0} N(k)!}$, where the factors $N(k)!$ in the denominator is the number of identical ways you can place the same balls in the same order into $N(k)$ boxes of size $k$. However, in the CBB-model one ball is assigned to each box to start with. This means that the distinguishable number of ways you can distribute the remaining $k-1$ balls into a box which already contains one ball, is $k$ times less the number of ways to put $k$ balls into the same box. Thus the total number of distinguishable ways you can distribute balls into the boxes is for the CBB-model reduced by a combinatorial factor $\Pi_{k=1} k^{N(k)}$ (more details in Supporting Information Text S1).

The most unbiased estimate of $\Omega$ corresponds to the maximum and this maximum corresponds to one particular distribution $N(k)$. This is completely equivalent to the maximum entropy principle in statistical mechanics [18]: Once the statistical states are identified, the most unbiased system corresponds to the maximum of such states or equivalently to the maximum of the entropy $S = \ln\Omega$. This maximum is achieved when all statistical states are equally probable. The last statement is the counterpart of the postulate of a priory equal probabilities is statistical mechanics [18]. From an information perspective the maximum of $S$ gives a measure of the maximum information which can be contained in the system [18]. For the CBB-model two statistical states are equal provided that there exists a one to one mapping between boxes containing the same balls and furthermore that for each such mapped pair of boxes the time order in which the balls arrived to the boxes differ by at most a cyclic permutation. The point is that this degeneracy of the statistical states is enforced by the constraint that the boxes always contain at least one ball: *The statistical state of the CBB-model and its network counterpart contain a k-degeneracy.*

The practical issue is to determine the distribution $N(k)$ which maximizes $\Omega[N(k)]$ subject to the appropriate constraints. One way is to find an update algorithm which picks the statistical states with equal probability, since such an algorithm automatically yields $N(k)$. For the CBB-model the obvious algorithm is as follows: *Pick two balls randomly and then move one to the same box as the other. If the attempted move involves emptying a box you try again with two new randomly picked balls.* If you start with a random distribution with at least one ball in each box, then the restriction of "always at least one ball in all boxes" is in this way implemented without any additional bias. The resulting distribution, $N(k)$, is shown in Fig. 2(a). An alternative is to directly find $N(k)$ using a mathematical standard method called variational calculus (see supporting information Text S1). Both alternatives give the same as result (see Fig. 2(a)). The advantage with the algorithm method is that it directly gives a complete characterization of the unbiased situation: the average $\langle N(k)\rangle$ which maximizes $S(\langle N(k)\rangle)$ and at the same time maximizes the noise $\Delta = \sqrt{\langle N(k)^2\rangle - \langle N(k)\rangle^2}$ [19]. As will be shown below this noise, or equivalently the spread of the data, provides important characteristics of the metabolic networks.

The definition of the statistical state is a direct consequence of the maximum number of distinguishable ways you can distribute the balls in the boxes: The statistical states, when picked with equal probability, give the global maximum of the entropy. However, our requirement is that the *relevant states* are picked with equal probability: in our case the rankings of the balls in the boxes is relevant. As emphasized above, this is a direct consequence of the sequential element implicit in the natural selection process and which imposes a time-order ranking on the balls in the boxes. *The crucial assumption is that the blind watchmaker network is random with respect to these relevant states.* The key observation is that many such relevant states correspond to the same statistical state. Thus unbiased, or equivalently random, with respect to the relevant states inevitably means biased with respect to the statistical states. Furthermore, biased with respect to the statistical states means smaller entropy $S$. What bias is imposed on the statistical states? In order to see this we note that the entropy $\tilde{S}$ is related to the probability $\tilde{p}$ of obtaining a new statistical state when choosing a new relevant state: $\tilde{S} = \ln\tilde{p}\Omega$. When the relevant states and statistical states are identical, this reduces to $\tilde{p} = 1$ and $S = \ln\Omega$. If several relevant states correspond to the same statistical state then $\tilde{p} < 1$ and $\tilde{S} < S$. In the present case $\tilde{p} = 1/\Pi_{k=1} k^{N(k)}$ because $\Pi_{k=1} k^{N(k)}$ is the number of different time-orders which gives the same statistical state. Thus $\tilde{S} = \ln\Omega/\Pi_{k=1} k^{N(k)} = \ln\Omega - S_a$ where $S_a = \sum_{k=1} N(k)\ln k$. Consequently, the most unbiased situation in terms of the relevant states corresponds to the $N(k)$ which maximizes $\tilde{S}$.

The corresponding least biased algorithm which achieves this maximization goes as follows:

1) pick two boxes (nodes) A and B randomly with probability $p \sim k^2$
2) pick a random ball in A and move to B.
3) If the attempted move is forbidden by a constraint choose another ball in A. Repeat until one ball is moved. Then choose two new boxes (nodes)
4) If no ball can be moved from A, choose two new boxes (nodes).

The important point here is that this algorithm incorporates the constraints in the most unbiased way and consequently yields the optimal $N(k)$. Figure 2(b) illustrates that the algorithm solutions which have the functional form $N(k) \sim \exp(-bk)/k^2$ (see Supporting Information in Text S1). Note that the distribution has an exponential decay for smaller values of average degree $\langle k\rangle = M/N$ but becomes very power law like for larger $\langle k\rangle$.

In order to find the variational solution for the network case, one needs to introduce the network constraints. These constraints can be directly implemented into the corresponding algorithm method: Moves which violates the network constraint are discarded.(see Supporting Information in Text S1) When the network constraints are included the distribution follows a power law $\langle N(k)\rangle \sim k^{-\gamma}$ over a large range with $\gamma > 2$, as illustrated in Fig. 2(c). An important point to notice is that average distribution $\langle N(k)\rangle$ is different from the individual network configurations $N(k)$. The striking difference is that a single network configurations always contains large split-off nodes, as illustrated in Fig. 2(d). *These split-off nodes constitute an essential characteristic of the network.*

## Discussion

How does the blind watchmakers random network compare to real networks? Fig. 3 illustrates this for the case of metabolic networks: Fig. 3(a) shows the average distribution obtained from 107 such networks and Fig. 3(b) the metabolic network for the E. Coli. bacteria (the data is taken from Ref. [14][15]). In Fig. 3(c) and 3(d) these networks are compared to the corresponding blind watchmaker network: the only information contained in this latter network is the number of nodes and links and the number of networks involved in the average. We again stress that both the shape of the distribution *and* the spread of the data are important characteristics of a real network. As apparent from Fig. 3(c) the overlap between the two data-sets is striking. From this we conclude that metabolic networks are to large extent blind watchmaker networks. Figure 3(d) compare a single metabolic network (E. Coli.) with the corresponding random network. A particular feature in this comparison is the large split off nodes. In case of E. Coli. there are 6 nodes with more than 100 links and for the corresponding blind watchmaker there are on the average 5.3±1.5.

Even if the blind watchmaker network explains the overall feature such as the fat tails of biological networks like the metabolic ones, there are of course differences. One such difference is the low number of nodes with only one link in case of metabolic networks. Such systematic deviations signal additional constraints in the real network. Whenever such constraints are present the entropy of the distribution is *lower* than for the blind watchmaker network. To illustrate this we in Fig. 3(e) and 3(f) compare the real networks with the corresponding blind watchmaker network *including* the least biased constraint which adjusts the number of single-link nodes (details in Supporting Information Text S1). Now the agreement is extraordinary considering the fact that only one data point, the number of single-link nodes at the very beginning of the distribution, has been adjusted. It clearly demonstrates that it is rather the *deviations* from the blind watchmaker network which needs to be explained (like in the present case the smaller number of single -link nodes): These deviations contain information about the system in addition to the less system specific information represented by the blind watchmaker network.

In the present work we focus on the different possible states a network can be found in. These network states distinguishes between the time order in which a node is connected to its neigbours. No a priory assumption of any growth mechanism or evolution process is made. We introduce the concept of *random with respect to the network states* and call the corresponding network *the blind watchmaker network*. It is found that the blind watchmaker network is scale-free and that metabolic networks to large extent are blind watchmaker networks. This means that the evolutionary path of the cell-metabolism, when projected onto a metabolic network representation, is statistically random with respect to a complete set of network states. This randomness emanates from the inherent randomness (the blind element) in Darwinian evolution and suggests that natural selection has had no or little effect on the network node degree distribution of metabolic networks.

Can these conclusions really be drawn from the agreement between the node degree distribution, the split off nodes and the stochastic spread of the data when comparing the actual data for metabolic networks and the blind watchmaker model network? In our opinion they can be drawn: The key is the quality of the agreement in relation to the number of free adjustable parameters. Earlier attempts to reproduce the degree distribution of metabolic networks usually starts from some assumption about the actual evolutionary path and has to our knowledge not been able to reproduce the data in the way demonstrated here [13]. So why does the evolution of metabolic networks choose this particular stochastic node degree distribution displayed by the blind watchmaker model network? According to us, the answer is that this degree distribution is *neutral* with respect to natural selection and hence, in this sense, has not been chosen at all. The notion that the node degree distribution for metabolic networks could be neutral with respect to natural selection has also been suggested on the bases of a comparison between the node degree distribution in atmospheric chemical reaction networks (no natural selection) and metabolic networks [20][21].

## Materials and Methods

The theoretical framework used in the analysis is classical statistical mechanics, both in its conventional variational form described in Text S1 and in its stochastical algorithm form described in Results.

The cell- metabolisms data represented by the metabolic networks are taken from Ma and Zeng in Ref. [14] and [15].

## Supporting Information

### Text S1

Found at: doi:10.1371/journal.pone.0001690.s001 (0.05 MB PDF)

## Author Contributions

Conceived and designed the experiments: SB PM. Performed the experiments: SB PM. Analyzed the data: SB PM. Contributed reagents/materials/analysis tools: SB PM. Wrote the paper: SB PM.